\documentclass[12pt]{iopart}
\usepackage{graphics}
\usepackage{epsfig}

\begin{document}

\title{Method for Generating
Additive Shape Invariant Potentials from an Euler Equation}
\author{Jonathan Bougie}
 \ead{$\!$jbougie@luc.edu, $\dag$agangop@luc.edu, $\ddag$jmallow@luc.edu }   
 \address{Loyola University Chicago, Department of Physics, Chicago,
IL 60660}
\author{Asim Gangopadhyaya$^{\dag}$}
 \address{Loyola University Chicago, Department of Physics, Chicago,
IL 60660}
\author{Jeffry V. Mallow$^{\ddag}$}
 \address{Loyola University Chicago, Department of Physics, Chicago,
IL 60660}
\date{\today}

\begin{abstract}
In the supersymmetric quantum mechanics formalism, the shape invariance condition provides a sufficient constraint to make a quantum mechanical problem solvable; i.e., we can determine its eigenvalues and eigenfunctions algebraically. Since shape invariance relates superpotentials and their derivatives at two different values of the parameter $a$, it is a non-local condition in the coordinate-parameter $(x, a)$ space. We transform the shape invariance condition for additive shape invariant superpotentials into two local partial differential equations.  One of these equations is equivalent to the one-dimensional Euler equation expressing momentum conservation for inviscid fluid flow. The second equation provides the constraint that helps us determine unique solutions.   We solve these equations to generate the set of all known $\hbar$-independent shape invariant superpotentials and show that there are no others.  We then develop an algorithm for generating additive shape invariant superpotentials including those that depend on $\hbar$ explicitly, and derive a new $\hbar$-dependent superpotential by expanding a Scarf superpotential.
\end{abstract}

\maketitle

\section {Introduction}

\subsection{Background}
Supersymmetric quantum mechanics (SUSYQM) provides an elegant and useful prescription for obtaining closed expressions for the energy eigenvalues and eigenfunctions of many one dimensional problems and three dimensional problems with spherical symmetry  \cite{Witten,CooperFreedman,Cooper-Khare-Sukhatme,Gangopadhyaya-Mallow-Rasinariu}.

In the SUSYQM formalism, first order differential operators $A^-$ and $A^+$ are
generalizations of the raising and lowering operators employed by Dirac for treating the harmonic oscillator \cite{Dirac}.  These ``ladder" operators $A^{\pm}(x,a_0) = \mp{\hbar}\frac{d}{dx} + W(x,a_0)$ use a superpotential $W$  to generate partner Hamiltonians  $H_- \equiv A^+ A^-$ and $H_+  \equiv A^- A^+$.  If these partner potentials are shape invariant, then the eigenspectra for both Hamiltonians can be derived algebraically without any prior information for either Hamiltonian.

Until recently, all known shape-invariant potentials could be generated from superpotentials with no explict dependence on $\hbar$ \cite{Cooper-Khare-Sukhatme,Gangopadhyaya-Mallow-Rasinariu}.  We classify such superpotentials as ``conventional."  However, a new class of shape-invariant potentials was discovered by Quesne \cite{Quesne} and expanded elsewhere \cite{Sasaki&Odake,odake}.  These potentials arise from an ``expanded" set of superpotentials that contain explicit $\hbar$-dependence.

In a recent publication \cite{Bougie}, we showed that the shape-invariance condition can be transformed into two local partial differential equations.  Solutions to these equations generate the set of all known conventional shape invariant superpotentials and allow no others in this category.  In addition, these equations provide an algorithm for generating $\hbar$-dependent potentials.  In this manuscript, we elaborate this method, proving the completeness of the set of ``conventional" superpotentials and generating a previously unknown ``expanded" superpotential.

\subsection {Supersymmetric Quantum Mechanics}

A quantum mechanical system in one spatial dimension $x$ described by a potential $V(x)$ can alternately be described by its ground state wavefunction $\psi_0$.  For simplicity of notation, we will use units where $2m=1$ throughout this manuscript.  We may also set the ground state energy to zero without changing the physics.  With these choices,the Schr\"odinger equation for the ground state wavefunction is 
\begin{eqnarray}  - \hbar^2 \psi^{''}_0 + V(x) \psi_0 =0~,
\end{eqnarray}
 where prime denotes differentiation with respect to $x$.
Thus, it follows that the potential can be written as
\begin{equation}V(x)=\hbar^2 \left({{\psi^{''}_0} \over {\psi_0}}\right).\label{potential}
\end{equation}

The SUSYQM formalism makes use of first order differential ``ladder" operators $A^-$ and $A^+$,
\begin{eqnarray}
   A^{\pm}(x,a_0) = \mp{\hbar}\frac{d}{dx} + W(x,a_0) \label{A}.
\end{eqnarray}
These ladder operators include a superpotential $W(x,a_0)$, which is a real function of coordinate $x$ and parameter $a_0$ (or a set of parameters).

Operators $A^\pm$ generate two supersymmetric partner Hamiltonians: $H_- \equiv A^+ A^-$ and $H_+  \equiv A^- A^+$. Hamiltonian $H_+$ is called the superpartner of $H_-$, and corresponding potentials $V_-$ and $V_+$ are given by
$V_\pm= W^2(x,a_0) \pm \hbar \frac{dW(x,a_0)}{dx}.$

\begin{figure}[h2]
\centering
    \epsfig{file=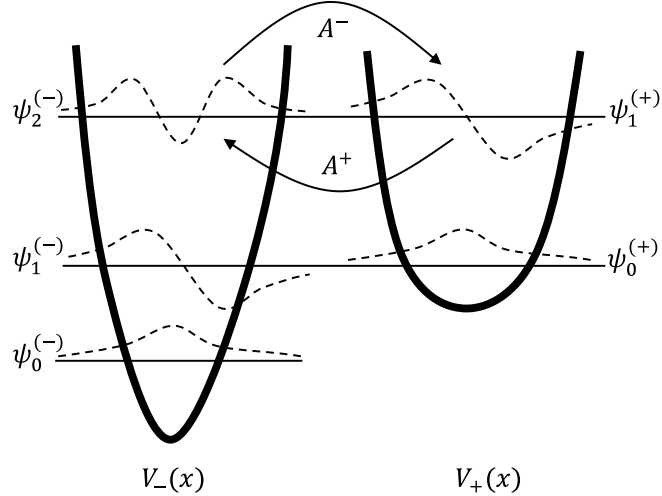, width=3.5in}
    \caption{Schematic illustrating the isospectrality of $H_+$ and $H_-$, showing the ground state and first two excited states of $H_-$ (corresponding to potential $V_-(x)$ sketched with a solid bold curve) in the left column, and the ground state and first excited state of $H_+$ (corresponding to potential $V_+(x)$, sketched as a solid bold curve) in the right column.  Sample energy levels are shown as horizontal lines, and a sketch of a sample wavefunction is overlaid with a dashed line for each energy level.  In general, $V_+$ and $V_-$ have different shapes, as do various $\psi^{+}$ and $\psi^{-}$.  However, the energy $E^+_1=E^-_0$, $E^+_2=E^-_1$, and $E^{+}_{n-1}\;=\;E^{-}_{n}$.  The operator $A^-$ acting on an eigenstate of $H_+$ will
yield an eigenstate of $H_-$, and $A^+$ acting on an eigenstate of $H_-$ will
yield an eigenstate of $H_+$.}
    \label{potentials2}
\end{figure}

Let us denote the eigenfunctions of $H_{\pm}$ that correspond to eigenvalues $E^{\pm}_n$, by $\psi^{(\pm)}_n$. For  positive integer $n$,
\begin{eqnarray}
H_{+} \left( A^-\psi^{(-)}_n \right) &=& A^-A^{+} \left( A^- \psi^{(-)}_n
\right)\nonumber\\
&=&A^- \left( A^{+}A^- \psi^{(-)}_n \right)\nonumber\\
& =&A^- H_{-} \left( \psi^{(-)}_n
\right)\nonumber\\
&=&E^{-}_n \left( A^- \psi^{(-)}_n \right).
\end{eqnarray}
Hence, with the exception of the the ground state which obeys $A^-\psi^{(-)}_0=0$, all excited states $\psi^{(-)}_n$ of $H_{-}$ have one-to-one correspondence with eigenstates  of $H_+$ with exactly the same energy: $\psi^{(+)}_{n-1} \propto A^- \psi^{(-)}_n$, where $E^{+}_{n-1}\;=\;E^{-}_{n}$ ($n=1,2, \cdots $), as illustrated in figure~\ref{potentials2}.
In other words, eigenstates of $H_+$ are iso-spectral with excited states of $H_-$.

Conversely,
$\psi^{(-)}_{n}  \propto A^+ \psi^{(+)}_{n-1}$. Thus, if the eigenvalues and the eigenfunctions of $H_{-}$ were known, one would automatically obtain the eigenvalues and the eigenfunctions of $H_{+}$ (and {\it vice versa}), which is in general a different Hamiltonian. However, unless we know one set of eigenstates {\it a priori}, this analysis is simply a mathematical curiosity.

\subsection {Shape Invariance}

Despite this limitation, for special cases in which these partner potentials obey the ``shape invariance" condition \cite{Infeld,gendenshtein}: \begin{eqnarray}
V_+(x,a_0)+g(a_0)=V_-(x,a_1)+g(a_1)~, \label{SIC}
\end{eqnarray} where parameter $a_1$ is a function of $a_0$, i.e., $a_1=f(a_0)$, the spectra for both Hamiltonians can be derived algebraically without any prior information for either Hamiltonian.  This is due to the existence of an underlying potential algebra  \cite{balantekin,asim1,asim2}.

Let us consider the special case where $V_-(x,a_0)$ is a shape invariant potential.  In this case, potentials $V_-$ and $V_+$ have the same $x$-dependence, and
the corresponding Hamiltonians $H_{+}(x,a_0)$ and $H_{-}(x,a_1)$ differ by $g(a_1)-g(a_0)$. Thus, their eigenfunctions are the same, and corresponding eigenvalues differ by $g(a_1)-g(a_0)$. In particular, they have a common ground state wavefunction, given by $\psi^{(+)}_{0}(x,a_0)={\psi^{(-)}_0(x,a_1)} \sim \exp\left( - \int^x_{x_0} W(x,a_1) dx \right)$, and the ground state energy of $H_{+}(x,a_0)$ is $g(a_1)-g(a_0)$, because the ground state energy of $H_{-}(x,a_1)$ is zero. Note that the parameter shift $a_0 \rightarrow a_1$ has an effect similar to that of a ladder operator: $\psi^{(-)}_{1}(x,a_0)  \sim A^+{(x,a_0)} ~ \psi^{(-)}_0(x,a_1)$.  Also  note that the ladder operators $A^-$ and $A^+$, like $H^\pm$, are dependent on parameters $a_n$.

The first excited state of $H_-(x,a_0)$ is given to within normalization by $A^+{(x,a_0)} \psi^{(-)}_0(x,a_1)$ and the corresponding eigenvalue is $g(a_1)-g(a_0)$. By iterating this procedure, the $(n+1)$-th excited state is given by
\begin{equation}
\psi^{(-)}_{n+1}(x,a_0) \sim
A^+{(a_0)} ~ A^+{(a_1)}  \cdots  A^+{(a_n)} ~ \psi^{(-)}_0(x,a_n)\;,
\label{wavefunction_sic}
\end{equation}
 and
corresponding eigenvalues are given by
\begin{equation}
E_0=0;
~~{\rm and}~~E_n^{(-)}=g(a_n)-g(a_0)~~{\rm
for}~n>0.
\label{energy_sic}
\end{equation}
 (To avoid notational complexity, we have suppressed the
$x$-dependence of operators $A(x,a_0)$ and $A^{+}(x,a_0)$.) Thus, for a shape invariant
potential, one can obtain the entire spectrum of $H_-$ itself by the algebraic methods of SUSYQM (and of course the same is true for $H_+$).

In this manuscript, we develop a method for finding shape-invariant superpotentials from a system of partial differential equations, and use this method to discover a new shape-invariant superpotential.  In Section II, we will show that for $\hbar$-independent (``conventional") superpotentials, the shape invariance condition can be converted into an infinite sequence of partial differential equations. In Section III, we solve these partial differential equations and systematically generate the complete set of conventional shape invariant potentials. In Section IV, we show a method for using these equations to find $\hbar$-dependent (``extended") superpotentials.  In Section V we use this method to find a new shape-invariant superpotential, and we present our conclusions in Section VI.

\section {Expressing the Shape Invariance Condition With Partial Differential Equations for Conventional Superpotentials}

As we have just seen, any partner potentials obeying (\ref{SIC}) can be solved algebraically.  Thus, discovering that a potential is shape-invariant yields much useful information.  Here we develop a method for finding shape-invariant potentials.

For shape invariant systems, the energy eigenvalues of $H_-(x,a_0)$ are given by $E^{(-)}_n(a_0) = g(a_n) -g(a_0)$, where $a_n \equiv f^n(a_0)$  indicates $f$ applied $n$-times to $a_0$ \cite{Cooper-Khare-Sukhatme,Gangopadhyaya-Mallow-Rasinariu}.  To avoid level crossing, $g(a)$ must be a monotonically increasing function of $a$; i.e., $\frac{\partial g}{\partial a}>0$.  In the case of ``additive" or ``translational" shape invariance, the parameters differ by an additive constant; i.e., $a_{i+1} = a_{i}+\hbar$.  While there are other forms of shape invariance such as multiplicative \cite{Barclay} and cyclic\cite{Cyclic}, most of the known exactly solvable superpotentials exhibit additive shape invariance \cite{CGK,Dutt}.   We thus restrict our analysis to additivel shape invariance.

Several groups found these potentials by imposing various ansatzes  \cite{CGK,asim2,asim3}.  In this manuscript, we derive these potentials {\it ab initio} as the solutions to a set of partial differential equations that must be satisfied for all additive
shape-invariant potentials.
Writing (\ref{SIC}) in terms of the superpotential, we obtain
\begin{eqnarray}
W^2(x,a_0) + \hbar \, \frac{d W(x,a_0)}{d x} + g(a_0)=
W^2(x,a_1)
- \hbar \,
\frac{d W(x,a_1)}{d x} + g(a_1).%
\label{SIC-SuperPotential}\end{eqnarray}
Note that (\ref{SIC-SuperPotential}) is a difference-differential equation; that is,  it relates the square of the superpotential $W$ and its spatial derivative computed at two different  parameter values: $(x,a_0\equiv a)$ and $(x,a_1 \equiv a+ \hbar)$.
This equation has also been studied for dynamical systems, where it is known as the infinite-dimensional dressing chain \cite{Shabat,spiridonov}.
Note that we have not needed to specify the value of  $\hbar$; therefore, this equation must hold for any value. Using this feature, we will transform (\ref{SIC-SuperPotential}) into a non-linear partial differential equation that is local in nature; i.e., all terms can be computed at the same point $(x,a)$.   This has the obvious advantage of mathematical familiarity (at least to physicists).  In addition, it provides a  systematic method for finding additive shape invariant potentials.

Important correspondences have been shown to exist between quantum mechanics and fluid mechanics \cite{Curtright}. SUSYQM is well known to have a deep connection with the KdV equation  \cite{KdV, sukumar,sukumar2, wang, kwong}, a nonlinear equation that describes waves in shallow water.
In this section we show that every additive shape invariant superpotential that does not depend explicitly on $\hbar$ corresponds to a solution of the Euler equation expressing momentum conservation for inviscid fluid flow in one spatial dimension.  We use this correspondence to develop a systematic method which 1) yields all such known $\hbar$-independent solvable potentials for SUSYQM and 2) shows that no others exist.  We will consider $\hbar$-dependent potentials in Section III.

Because of additive shape invariance, the dependence of $W$ on $a$ and $\hbar$ is through the linear combination $a+ \hbar$; therefore, the derivatives
of $W$ with respect to $a$ and $\hbar$ are related by: $ \frac{\partial W(x,a+ \hbar)}{\partial \hbar} = \,\frac{\partial W(x,a+ \hbar)}{\partial a}$.
Since (\ref{SIC-SuperPotential}) must hold for an arbitrary value of $\hbar$, if we assume that $W$ does not depend explicitly on $\hbar$, we can expand in powers of $\hbar$ and the coefficient of each power must separately vanish. Expanding the right hand side in powers of $\hbar$ yields
\begin{eqnarray}%
{\cal O}(\hbar)&\Rightarrow& W \, \frac{\partial W}{\partial a} - \frac{\partial W}{\partial x} + \frac12 \, \frac{d g(a)}{d a} = 0~,
\label{PDE1}\\
{\cal O}(\hbar^2)&\Rightarrow& \frac{\partial }{\partial a}\left( W \, \frac{\partial W}{\partial a} - \frac{\partial W}{\partial x} + \frac12 \, \frac{d g(a)}{d a} \right)= 0~,
\label{PDE2}\\
{\cal O}(\hbar^n)&\Rightarrow& \frac{\partial^{n}}{\partial a^{n-1}\partial x} ~W(x,a)= 0~, ~~~~~n\geq 3 ~.
\label{PDE3} \end{eqnarray}
Thus, all conventional additive shape invariant superpotentials are solutions of the above set of non-linear partial differential equations.  Although this represents an infinite set, note that if equations at ${\cal O}(\hbar)$ and ${\cal O}(\hbar^3)$ are satisfied, all others automatically follow.

Replacing $W$ by $-u$, $x$ by $t$, and $a$ by $x$
in (\ref{PDE1}), the equation then becomes:
\begin{eqnarray}%
\left(u(x,t) \,  \frac{\partial  }{\partial
x}\right) u(x,t)+ \frac{\partial u(x,t) }{\partial t}  = -\frac12
\frac{d g(x)}{d x}~.
\end{eqnarray}
This equation is equivalent to the equation for inviscid fluid flow in the absence of an external force on the bulk of the fluid:
\begin{equation}\frac{\partial {\bf u}\left({\bf x},t\right)}{\partial t}+u\left({\bf x},t\right)\cdot\nabla {\bf u}\left({\bf x},t\right) = -\frac{\nabla p\left({\bf x},t\right)}{\rho\left({\bf x},t\right)}\label{Euler}\end{equation}
in one spatial dimension with the correspondence $\frac{1}{\rho}\frac{d p}{d x} = \frac 12 \frac{d g}{d x}$, where ${\bf u}$ is the fluid velocity at location ${\bf x}$ and time $t$, $p$ is the pressure, and $\rho$ is the local fluid density. Equation~(\ref{Euler}) is one of the fundamental laws of fluid dynamics, and was first obtained by Euler
in 1755  \cite{Euler}.  Thus, all conventional shape invariant superpotentials form a set of solutions to the one-dimensional Euler equation.

It should be noted that (\ref{Euler}) is not solvable unless additional constraints are applied.  In fluid dynamics this equation is generally supplemented by the continuity equation expressing conservation of mass, along with an equation of state and/or the energy equation and boundary conditions. These additional constraints do not apply in SUSYQM.  Instead, (\ref{PDE3}) supplies the additional constraint which must be fulfilled to satisfy shape-invariance.  Thus, the set of solutions of (\ref{PDE1}) that also satisfy the constraint of (\ref{PDE3}) will define the complete set of conventional shape invariant superpotentials.

\section {Generating the complete set of conventional superpotentials}
\subsection {Solutions of Special Cases}

In this section, we show that the set of all possible conventional superpotentials are determined by six special cases.  To find this set of solutions, we note that (\ref{PDE3}) is satisfied for all $n\geq 3$ as long as
\begin{eqnarray}
\frac{\partial^{3}}{\partial a^{2}\partial x} ~W(x,a)= 0.\label{Eq1}
\end{eqnarray}
The general solution to Eq.~ (\ref{Eq1}) is
\begin{eqnarray}
W(x,a)=a\cdot {X}_1(x)+{X}_2(x)+u(a)~.\label{GeneralSolution}
\end{eqnarray}

Substituting this into (\ref{PDE1}) yields

\begin{eqnarray}\underbrace{\left(a\cdot {X}_1+{X}_2+u\right)}_W \underbrace{\left({X}_1+\frac{d u}{d a} \right)}_{\frac{\partial W}{\partial a}}-
\underbrace{\left(a\cdot \frac{d {X}_1}{d x}+\frac{d {X}_2}{d x}\right)}_{\frac{d W}{d x}}+~\frac12~~ \frac{d g}{d a}  = 0~~.
\end{eqnarray}
To systematically find all possible solutions, we define $u\, \frac{d u}{d a}+\frac12~\frac{d g}{d a}=-H(a)$, and then collect and label terms based on their dependence on ${X}_{1}$ and ${X}_{2}$ and their derivatives:
\begin{eqnarray}\fl
\underbrace{{X}_1\,{X}_2}_{\mbox{Term\#1}} +
\underbrace{\left(- \frac{d {X}_2}{d x}\right)}_{\mbox{Term\#2}}
+
\underbrace{a\, {X}_1^2}_{\mbox{Term\#3}}+
\underbrace{\left( - a \frac{d {X}_1}{d x}\right)}_{\mbox{Term\#4}}+
\underbrace{\frac{d u}{d a}\,{X}_2 }_{\mbox{Term\#5}}+
\underbrace{\left(u+a\frac{d u}{d a}\right){X}_1}_{\mbox{Term\#6}}=
H(a)~.\label{Eqterms}
\end{eqnarray}

We begin by considering special cases of (\ref{Eqterms}) where one or more of the terms ${X}_1(x)$, ${X}_2(x)$, or $u$ can be considered zero.  After considering these cases, we will show that all solutions to this equation can be reduced to one of these cases.
In the nomenclature that follows, lower case Greek letters denote $a$- and $x$-independent constants.

\subsubsection{Case 1: ${X}_{2}$ and $u$ are not constants, ${X}_{1}$ is constant.}
In this case, let ${X}_{1}=\mu$. Then our general form for $W$ becomes $W=\mu a + u(a) + {X}_{2}(x)$.  If we define $\tilde{u}\equiv u(a)+\mu a$, we get $W=\tilde{u}+{X}_{2}$.  So this case is equivalent to ${X}_{1}=0$.   Then terms 1, 3, 4, and 6 each become zero, and (\ref{Eqterms}) becomes $-\frac{d {X}_{2}}{dx}+\frac{du}{da}{X}_{2}= H(a)$.    Since ${X}_{2}$ must be  independent of $a$ but cannot be constant,$\frac{d {X}_{2}}{dx}\neq 0$.  Thus, this is possible only if $\frac{du}{da}$ and $H(a)$ are independent of $a$. This yields $u=\alpha a + \beta $,  $H(a)=\theta$. Therefore, $-\frac{d {X}_{2}}{dx}+\alpha{X}_{2}= \theta$, where $\alpha\neq 0$ since $u$ is not constant.  The solution to this differential equation is ${X}_{2}(x)=\frac{\theta}{\alpha}+\eta\, e^{\alpha x}$.  Therefore, $W=\alpha a +\beta +\frac{\theta}{\alpha}+\eta\, e^{\alpha x}$. Defining $\alpha = -1$, this yields $W=A-B e^{-x}$, where $A\equiv \beta -a-\theta$.  This is the Morse superpotential.

\subsubsection{Case 2: ${X}_{1}$ and $u$ are not constants, ${X}_{2}$ is constant.}
Following a similar procedure this case is equivalent to ${X}_{2}=0$.
 So $a X_1^2-a\frac{d\chi_1}{dx}+\left(u+a\frac{du}{da}\right) X_1=H(a)$.
Since $X_1$  cannot depend on $a$ but must contain $x$-dependence, the only way for $H(a)$ to be independent of $x$ is if $u+a\frac{du}{da}=\alpha a$, where $\alpha$ could be any constant.  Thus,
$u=\frac{\alpha a}{1}+\frac{\beta}{a}$, where $\alpha$ and $\beta$ could be any two constants, although they cannot both be zero.

Thus, $a X_2^2-a\frac{dX_1}{dx}+a\alpha X_1=H(a)$.  This is only possible if $H(a)=\gamma a$ for some constant $\gamma$.  So $X_1^2-\frac{dX_1}{dx}+\alpha X_1 = \gamma$.  This differential equation gives different solutions depending on the values of $\alpha$ and $\gamma$.

If $\alpha \neq 0$, then  $\gamma \neq 0$ yields the Rosen-Morse I superpotential, and $\gamma = 0$ yields the Eckart superpotential.

If $\alpha =0$, then $ \gamma=0$ yields Coulomb, $\gamma>0$ yields Rosen-Morse II, and $\gamma<0$ yields Rosen-Morse I.

Thus, the  Rosen-Morse I, Rosen-Morse II, Eckart, and Coulomb superpotentials are all solutions to Case 2, for different values of $\alpha$ and $\gamma$.

\subsubsection{Case 3: ${X}_{1}$ and ${X}_{2}$ are not constants, $u=\mu a + \nu$ (this includes the case where u is constant.)}
In this case we can define $\tilde{X}_{1} \equiv {X}_{1}+\mu$ and $\tilde{X}_{2} \equiv {X}_{2}+\nu$, which is equivalent to $u=0$.  For $u=0$, $X_1 X_2 - \frac{dX_2}{dx}+a\left(X_1^2-\frac{dX_1}{dx}\right)=H(a)$.

Since $X_1$ and $X_2$ are both independent of $a$, the coefficients of each power of $a$ must cancel separately.  So we are left with two coupled differential equations: $X_1^2-\frac{dX_1}{dx}=\alpha$, $X_1 X_2-\frac{dX_2}{dx}=\beta$.  Again, the solution varies depending on the values of $\alpha$ and $\beta$, which could be any constants.

If $\alpha=0$, the solution is the 3-D harmonic oscillator superpotential.  If $\alpha<0$, the solution is the Scarf I superpotential, and if $\alpha>0$, the equation is solved by either the Scarf II or Generalized P\"{o}schl-Teller superpotential.

Therefore, the Scarf I, Scarf II, 3-D oscillator, and Generalized P\"{o}schl-Teller superpotentials are all solutions of Case 3.

\subsubsection{Case 4: ${X}_{2}$ is not constant, ${X}_{1}$ and $u$ are constant.}
If ${X}_{1}\neq 0$, this is equivalent to $X_1=0$ and $u=\mu a + \nu$.  In such a case, this is equivalent to Case 1, and the solution is the Morse superpotential.  However, if ${X}_{1}=0$, this case is equivalent to $X_1=u=0$, in which case $-\frac{dX_2}{dx}=H(a)$.  Since $X_2$ is independent of $a$, $H(a)$ must be a constant.  This generates the one-dimensional harmonic oscillator.

\subsubsection{Case 5: ${X}_{1}$ is not constant, ${X}_{2}$ and $u$ are constant.}
In this case, $\alpha X_1 + a X_1^2-a\frac{dX_1}{dx}=H(a)$ for some constant $\alpha$.  Since $X_1$ is independent of $a$ but must depend on $x$, $\alpha=0$ and $H(a)=\beta a$ for some constant $\beta$.  Thus, $X_1^2-\frac{dX_1}{dx}=\beta$.
This yields special cases of Scarf I and Scarf II, and the centrifugal term of the Coulomb and 3-D oscillator, depending on whether $\beta$ is positive, negative, or zero.

\subsubsection{Case 6: ${X}_{1}$ is constant, ${X}_{2}$ is constant.}
In this case, the superpotential has no $x$-dependence, regardless of the value of $u$.  This is a trivial solution corresponding to a flat potential, and we disregard it.

Thus far, we have considered all of the special cases that are equivalent to $u=0$, $X_1=0$, or $X_2=0$.  These special cases generate all known conventional additive shape-invariant superpotentials \cite{Cooper-Khare-Sukhatme,Dutt,Gangopadhyaya-Mallow-Rasinariu}. We have listed them all in Table \ref{Table1}.

\begin{table} [htb]
\begin{center}
\begin{tabular}{||l|l|l||}
\hline Name  &  superpotential   & Special \\
&  & Cases \\ \hline
Harmonic Oscillator &  $\frac12 \omega x$ & $X_1=u=0$  \\
  \hline
Coulomb   &  $\frac{e^2}{2(\ell+1)} - \frac{\ell+1}{r}$& $X_2=0$  \\
    \hline
3-D oscillator  & $\frac12 \omega r - \frac{\ell+1}{r}$ &  $u=0$\\
 \hline
Morse &$A-Be^{-x}$ &  $X_1=0$ \\
 \hline
Rosen-Morse I &$-A\cot x - \frac{B}{A}$ & $X_2=0$  \\
 \hline
Rosen-Morse II &$A\tanh x + \frac{B}{A}$ & $X_2=0$   \\
 \hline
Eckart &$-A\coth x + \frac{B}{A}$ & $X_2=0$  \\
 \hline
{Scarf I} &$ A\tan x - B {\rm sec}\,x$ & $u=0$\\
 \hline
{Scarf II}  &$A\tanh x + B {\rm sech}\,x$& $u=0$\\
 \hline
Gen. P\"oschl-Teller &$A\coth x- B {\rm cosech}\,x $ & $u=0$ \\
 \hline
\end{tabular}
\caption{The complete family of additive shape-invariant superpotentials.}\label{Table1}

\end{center}
\end{table}

\subsection{Proof That the List of Conventional Superpotentials is Complete}

Now that we have these special cases, we can systematically obtain all possible solutions.  $H(a)$ is independent of $x$.  Therefore, when any solution is substituted into (\ref{Eqterms}), it will yield an $x$-independent sum of terms 1-6.  There are many ways in which these terms could add up to a term independent of $x$. As a first step, we begin with the simplest possibility, in which each term is individually independent of $x$.

Under this assumption, term 3 states that ${X}_{1}$ must be a constant, independent of $x$.  In addition, term 1 dictates ${X}_{1} {X}_{2}$ must be constant as well.  These two statements can only be true if ${X}_{2}$ and ${X}_{1}$ are constant separately; this reduces to the trivial solution of case 6.

Therefore, assuming that each term is separately independent of $x$ yields only the trivial solution.  However, there is also the possibility that some of the terms depend on $x$, but when added to other terms, the $x$-dependence cancels to yield a sum that is independent of $x$.   If a group of $n$-terms taken together produce an $x$-independent sum, and if no smaller subset of these terms add up to a sum independent of $x$, we call this group of $n$-terms ``irreducibly independent of $x$".

As an example, we check whether there are any solutions for the six-term irreducible set in which the sum of all six terms in (\ref{Eqterms}) is independent of $x$, but in which the sum of any subset of terms would depend on $x$.

To check this possibility, we note that the first two terms are independent of $a$, while terms 3 and 4 are linear in $a$.  We do not know {\it a priori} the functional form of $u$.   Since term 5 contains $\frac{du}{da}$, it could include terms independent of $a$, terms linear in $a$, and/or other forms of $a$-dependence.  Similarly, we do not know the functional form of $u+a\frac{du}{da}$.  Therefore, we define functions $f_1(a)$ and $f_2(a)$ such that $u+a\frac{du}{da}=\alpha_1+\beta_1 a +f_1(a)$ and  $\frac{du}{da}=\alpha_2+\beta_2 a +f_2(a)$, where we do not know the functional form of $f_1(a)$ and  $f_2(a)$, except that they contain no constant terms or terms linear in $a$.

With this definition, term 6 becomes $\alpha_1 X_1 + a \beta_1 X_1 + f_1(a) X_1$ and term 5 becomes $\alpha_2 X_2 + a \beta_2 X_2 + f_2(a) X_2$ Since $f_1(a)$ and $f_2(a)$ contain no constant terms or terms proportional to $a$, the $x$ dependence of the term $f_1(a) X_1$ cannot be cancelled by terms 1-4, and neither can the $x$-dependence of   $f_2(a) X_2$.  Therefore, we conclude that $f_1(a) X_1 + f_2(a) X_2 = \mu$.

This leaves only two possibilities.  First we consider the possibility that $f_1(a)=f_2(a)=0$.   In this case, $\frac{du}{da}=\alpha_2 + \beta_2 a$, so $u=\alpha_2 a + \frac{\beta_2 a^2}{2}+\gamma$.  Plugging this solution into (\ref{Eqterms}), term 6 then becomes $\left(\frac{3\beta_2}{2}a^2+2\alpha_2+\gamma\right)X_1$.  Since $X_1$ is independent of $a$ and none of the other terms are proportional to $a^2$, $\frac{3\beta_2}{2}X_1$ must be independent of $x$.  So either $X_1$ is a constant, or $\beta_2=0$, in which case $u$ is linear in $a$; either possibility represents a special case.

On the other hand, if $f_1(a) \neq 0$, then either $X_1$ is a constant (which is a special case), or the $x$-dependence of $f_1(a) X_1$ must be cancelled by the $x$-dependence of $ f_2(a) X_2$.  Since $X_1$ and $X_2$ cannot depend on $a$, this is only possible if ${X}_{2}+\mu {X}_{1}=\nu$ for some constants $\mu$ and $\nu$.
However, if this is the case, then $W=aX_1+X_2+u=aX_1-\mu X_1 + \nu + u$.  Since the zero of $a$ can be shifted by defining ${\tilde a}=a-\mu$, and we can define ${\tilde u}=u+\nu$, this is equivalent to the case $W={\tilde a}X_1 + {\tilde u}$.  So this particular case can be reduced to the special case of $X_2=0$ and does not yield any new solutions.
Thus, no new solutions can be found by assuming the full six-term set is irreducibly independent of $x$; any solution found under this assumption can be reduced to one of the special cases.

Since each term cannot be separately independent of $x$ and the full six-term set cannot be irreducibly independent of $x$, the only possibility that could produce solutions not covered under one of the special cases would be if two or more irreducibly independent sets combine to produce a solution.   If, for example, term 5 depends on $x$ and term 6 depends on $x$, but the sum of these two terms is $x$-independent, then we consider the set of terms $\{5,6\}$ to be a two-term set that is irreducibly independent of $x$.  To continue this example, if the four sets $\{1\}$, $\{ 2,3 \}$,
$\{ 4 \}$, and $\{5, 6\}$ were each irreducibly independent of $x$, then the entire left side of (\ref{Eqterms}) would be independent of $x$, and this could possibly produce a solution not covered under the special cases.  We now proceed to examine all possible combinations and show that no new solutions are, in fact, produced.

To examine all possibilities, we begin by examining all possible one-, two-, and three- term sets, and see what restrictions are imposed on solutions in each case.  For example, if term 1 is individually independent of $x$, then $X_1 X_2=\alpha$ for some constant $\alpha$, which implies that $X_1 = \alpha / X_2$.  Table \ref{1termsets} shows the consequences required for each possible one-term set to be independent of $x$.

\begin{table} [htb]
\begin{center}
\begin{tabular}{|c|c|c|}
\hline Single term& Requirements for   &  Compatible with \\
set & $x$-independence  &  a new solution? \\
\hline
$\{1\}$ &$X_1 = \alpha/X_2$ & If $\alpha \neq 0$\\
\hline
$\{2\}$ &$X_2 = \alpha x + \beta$ & If $\alpha \neq 0$\\
\hline
$\{3\}$ &$X_1 = \alpha$ & No\\
\hline
$\{4\}$ &$X_1 = \alpha x+\beta$ & If $\alpha \neq 0$\\
\hline
$\{5\}$ &$X_2 = \alpha$ & No\\
\hline
$\{6\}$ &$u=\alpha/a$ & If $\alpha \neq 0$\\
\hline
\end{tabular}
\caption{Assuming that a single term is independent of $x$ results in restrictions on the possible solutions.  The first column lists each possible single term set.  The middle column of each row shows what consequences are necessary for that set to be independent of $x$.  Throughout this section, lower-case greek letters indicate constants that are independent of both $a$ and $x$. Finally, if this requirement is compatible with solutions not included as one of the special cases, restrictions on such a solution are shown in the third column of that row.  The third column states ``No" if the requirement is equivalent to a special case and thus no new solutions are allowed.}\label{1termsets}
\end{center}
\end{table}

We now proceed to examine two-term sets.  Let us take the example mentioned above in which $\{5,6\}$ is a two-term set that is irreducibly independent of $x$ (that is, term 5 depends on $x$ and term 6 depends on $x$, but the sum of these two terms is $x$-independent).  Let us consider this example further to see if it leads to any new solutions.  In this example, $\frac{du}{da}X_2+\left(u+a\frac{du}{da}\right)X_1$ must be independent of $x$.  Since $X_1$ and $X_2$ must each depend on $x$ and $\frac{du}{da}$ must depend on $a$ (or this would reduce to a special case), the only way this is possible is if the $x$-dependence of $\frac{du}{da}X_2$ is cancelled by the $x$-dependence of $\left(u+a\frac{du}{da}\right)X_1$.  For this to be true, the $a$-dependence of $\frac{du}{da}$ must differ from the $a$-dependence of $u+a\frac{du}{da}$ by only a multiplicative constant.  Thus, we conclude that
$X_2+\alpha X_1 = \beta$ for some constants $\alpha$ and $\beta$. As we did when considering the full six-term set, we can absorb ${X}_{2}$ into ${X}_{1}$ by shifting the zero of $a$.
So this particular case can be reduced to the special case of $X_2=0$ and does not yield any new solutions.

However, not all two-term sets can be reduced in this way.  As an example, we consider the set $\{ 1,2\}$.  For this set to be irreducibly independent of $x$, $X_1 X_2$ and $\frac{dX_2}{dx}$ must each depend on $x$.  However, $X_1 X_2-\frac{dX_2}{dx}$ must be independent of $x$.  Since $X_1$ and $X_2$ do not depend on $a$, the only way for this to be possible is if $X_1 X_2-\frac{dX_2}{dx}=\alpha$ for some constant $\alpha$.  By itself, this requirement could allow many solutions.   Therefore, we will have to check whether this set can couple with the remaining terms 3, 4, 5, and 6 in such a way as to produce new solutions that are not compatible with the special cases.

In Table \ref{2termsets}, we display the consequences required for each possible two-term set to be irreducibly independent of $x$.  From this table, it is clear that there are several two-term sets that can be irreducibly independent of $x$.  However, in order to satisfy  (\ref{Eqterms}), these sets must be compatible with solutions that allow the remaining four terms to be independent of $x$.  We now check whether there are any combinations of one- and two- term sets that can satisy  (\ref{Eqterms}).  Once we have completed this, we will consider cases involving three-, four-, and five- term irreducible sets.

\begin{table} [htb]
\begin{center}
\begin{tabular}{|c|c|c|}
\hline Two-term& Requirements for &  Compatible with \\
set& irreducible $x$-independence &  a new solution? \\
\hline
$\{1,2\}$ & $X_1 X_2 -\frac{dX_2}{dx}=\alpha$ & For any $\alpha$\\
\hline
$\{1,3\}$ & Contradiction: must reduce & No\\
\hline
$\{1,4\}$ & Contradiction: must reduce & No\\
\hline
$\{1,5\}$ & $u=\mu a + \nu$ & No\\
\hline
$\{1,6\}$ & $u=\frac{\alpha}{a}+\gamma$; & If $\alpha \neq 0$; $\beta\neq 0$\\
&$X_1=\frac{\beta}{X_2+\alpha}$&\\
\hline
$\{2,3\}$ & Contradiction: must reduce & No\\
\hline
$\{2,4\}$ & Contradiction: must reduce & No\\
\hline
$\{2,5\}$ & $u=\mu a + \nu$ & No\\
\hline
$\{2,6\}$ & $u=\frac{\alpha}{a}+\gamma$;  & If $\alpha \neq 0$\\
& $-\frac{dX_2}{dx}+\alpha X_1=\beta$&\\
\hline
$\{3,4\}$ & $X_1^2-\frac{dX_1}{dx}=\alpha$ & For any $\alpha$\\
\hline
$\{3,5\}$ & $u=\frac{\alpha a^2}{2}+\gamma$; & If $\alpha \neq 0$\\
&$X_1^2+\alpha X_2 = \beta$ &\\
\hline
$\{3,6\}$ & Contradiction: must reduce & No\\
\hline
$\{4,5\}$ & $u=\frac{\alpha a^2}{2}+\gamma$; & If $\alpha \neq 0$\\
&$\frac{dX_1}{dx}=\alpha X_2 + \beta$ &\\
\hline
$\{4,6\}$ & $u=\frac{\alpha}{a}+\gamma$;  & If $\alpha \neq 0$; $\beta \neq 0$\\
&$X_1=\beta e^{\alpha x} + \mu$&\\
\hline
$\{5,6\}$ & $X_2=\alpha X_1 + \beta$ & No \\
\hline
\end{tabular}
\caption{Assuming that a two-term set is irreducibly independent of $x$ results in restrictions on the possible solutions.  The first column lists each possible two-term set.  The middle column of each row shows what consequences are necessary for that set to be irreducibly independent of $x$.  The middle column states ``Contradiction: must reduce" if the assumption that the two-term set is irreducibly independent of $x$ leads to the contradictory conclusion that the set must be reducible.  Finally, if this requirement is compatible with solutions not included as one of the special cases, restrictions on such a solution are shown in the third column of that row.  The third column states ``No" if no new solutions are allowed, either because the assumption of irreducibility leads to a contradiction or because the requirement on solutions is equivalent to a special case.}\label{2termsets}
\end{center}
\end{table}

Combining the results from Table \ref{1termsets} and  Table \ref{2termsets} allow us to immediately eliminate many possibilities.  For instance, in the example listed above, the set $\{ 1,2\}$ is irreducibly independent if $X_1 X_2-\frac{dX_2}{dx}=\alpha$ for some constant $\alpha$.  However, in order to satisfy (\ref{Eqterms}), the remaining terms must combine in such a way that the combination of terms 3, 4, 5, and 6 are independent of $x$ as well.  We first note from Table \ref{1termsets} that term 3 cannot be independent of $x$ by itself, and neither can term 5.  From Table \ref{2termsets}, we note that $\{3,6 \}$ cannot be irreducibly independent of $x$, and the set $\{5,6 \}$ is equivalent to a special case.

Therefore, there are only two possible combinations of one- and two- term sets involving the irreducible set $\{ 1,2\}$ that could satisfy (\ref{Eqterms}) and lead to a new solution.  The first possibility is that set $\{1,2\}$ is irreducibly independent of $x$, set  $\{3,5\}$ is irreducibly independent of $x$, and set  $\{4,6\}$ is irreducibly independent of $x$.  However, from Table \ref{2termsets}, $\{3,5\}$ requires $u=\frac{\alpha a^2}{2}+\gamma$ with $\alpha \neq 0$.  On the other hand,  $\{4,6\}$ requires $u=\frac{\alpha}{a}+\gamma$, so these two terms are incompatible.  The only other possibility is that set $\{1,2\}$ is irreducibly independent of $x$, set  $\{3,5\}$ is irreducibly independent of $x$, and sets  $\{4\}$ and $\{6\}$ are each separately independent of $x$.  However, once again the requirement $u=\alpha/a$ for term 6 to be $x$-independent is incompatible with the requirement $u=\frac{\alpha a^2}{2}+\gamma$ with $\alpha \neq 0$ for $\{3,5\}$.  Therefore, there can be no possible new solutions resulting from combinations of one- and two- term sets involving the irreducible set $\{ 1,2\}$.

Comparing Table \ref{1termsets} and  Table \ref{2termsets}, every
possible combination of one- and two- term sets leads to one of the following three results: a) one of the terms is directly equivalent to a special case (such as any combination involving the single-term set $\{3\}$); b) one of the terms leads to a contradiction in which a term assumed to be irreducible must be reducible (such as any combination involving the two-term set $\{1,3\}$); or c) two elements in the combination require different functional forms of $u$, leading to the impossibility of a common solution (such as any combination involving the sets $\{3,5\}$ and $\{4,6\}$).  Therefore, we conclude that no new solutions can be found from combinations of one- and two- term sets.

However, there is still the possibility that there are solutions provided by combinations involving three-, four-, or five-term irreducible sets.  We now test these possibilities, begining with three-term sets. In Table \ref{3termsets}, we display the consequences required for each possible three-term set to be irreducibly independent of $x$.

\begin{table} [htb]
\begin{center}
\begin{tabular}{|c|c|c|}
\hline Three-term& Requirements for &  Compatible with \\
set& irreducible $x$-independence &  a new solution? \\
\hline
$\{1,2,3\}$ &  Contradiction: must reduce & No\\
\hline
$\{1,2,4\}$ & Contradiction: must reduce & No\\
\hline
$\{1,2,5\}$ & $u=\mu a + \nu$ & No\\
\hline
$\{1,2,6\}$ & $u=\frac{\alpha}{a}+\gamma$;  & If $\alpha \neq 0$\\
&$X_1 X_2-\frac{dX_2}{dx}+\alpha X_1=\mu$&\\
\hline
$\{1,3,4\}$ & Contradiction: must reduce & No\\
\hline
$\{1,3,5\}$ & Contradiction: must reduce & No\\
\hline
$\{1,3,6\}$ & Contradiction: must reduce & No\\
\hline
$\{1,4,5\}$ & $u=\frac{\alpha a^2}{2}+\beta a + \gamma$; & If $\alpha \neq 0$; $\beta \neq 0$\\
&$X_2=\frac{\nu}{X_1+\beta}$;&\\
& $-\frac{dX_1}{dx}+\frac{\alpha \nu}{X_1+\beta}=\mu$ &\\
\hline
$\{1,4,6\}$ & $u=\frac{\alpha a}{2}+\beta +\frac{\gamma}{a}$  &$\alpha \neq 0$; $\beta \neq 0$\\
&$X_1=\frac{\nu}{X_2+\beta}$&\\
&$X_1=\mu e^{\alpha x} + \eta$&\\
\hline
$\{1,5,6\}$ & $X_2=\alpha X_1 + \beta$ & No\\
\hline
$\{2,3,4\}$ & Contradiction: must reduce & No\\
\hline
$\{2,3,5\}$ & $u=\frac{\alpha a^2}{2}+\beta a+\gamma$; & If $\alpha \neq 0$; $\beta \neq 0$;\\
&$X_2=\mu e^{\beta x}+\nu$;&$\mu \neq 0$\\
&$X_1^2+\alpha X_2 =\mu$&\\
\hline
$\{2,3,6\}$ & Contradiction: must reduce & No\\
\hline
$\{2,4,5\}$ & $u=\frac{\alpha a^2}{2}+\beta a + \gamma$; & If $\alpha \neq 0$; $\beta \neq 0$\\
&$X_2=\mu e^{\beta x}+\nu$;&$\mu \neq 0$\\
&$-\frac{dX_1}{dx}+\alpha X_2=\eta$&\\
\hline
$\{2,4,6\}$ & $u=\frac{\alpha a}{2}+\beta+\frac{\gamma}{a}$; & If $\alpha \neq 0$; $\beta \neq 0$\\
&$X_2=\mu e^{\beta x}+\nu$;&$\mu \neq 0$\\
&$-\frac{dX_2}{dx}+\beta X_1=\eta$&\\
\hline
$\{2,5,6\}$ & $X_2=\alpha X_1 + \beta$ & No \\
\hline
$\{3,4,5\}$ & Contradiction: must reduce & No\\
\hline
$\{3,4,6\}$ & $u=\frac{\alpha a -\gamma}{2}+\frac{\gamma}{a}$ &  If $\alpha \neq 0$\\
&$X_1^2-\frac{dX_1}{dx}+\alpha X_1=\mu$&\\
\hline
$\{3,5,6\}$ & Contradiction: must reduce &  No\\
\hline
$\{4,5,6\}$ & $X_2=\alpha X_1 + \beta$ & No \\
\hline
\end{tabular}
\caption{Assuming that a three-term set is irreducibly independent of $x$ results in restrictions on the possible solutions.  The first column lists each possible three-term set.  The middle column shows the consequences necessary for that set to be irreducibly independent of $x$.  If this requirement is compatible with solutions not included as one of the special cases, restrictions on such a solution are shown in the third column.}\label{3termsets}
\end{center}
\end{table}

Following the same procedure as for two-term sets, we see that  once again all combinations produce one of the following results: a) one of the terms is directly equivalent to a special case; b) one of the terms leads to a contradiction in which a term assumed to be irreducible must be reducible; or c) two elements in the combination require different functional forms of $u$, leading to the impossibility of a common solution.

We now turn to the case of four-term irreducible sets.  In this case, rather than calculating the restrictions on all possible four-term sets, we use the fact that each four-term set must combine with either a two-term set or a pair of single-term sets.  We can use this fact to eliminate many possibilities.  For instance, in order for set $\{2,4,5,6\}$ to yield new solutions, it must combine either with the two-term set $\{1,3\}$ or with the pair of single term sets $\{1\}$ and $\{3\}$.  Since neither possibility can yield new solutions, we can eliminate the four-term set  set $\{2,4,5,6\}$ from consideration.

Of the remaining four-term sets, many more can be considered by assuming the requirements from the complementary one- or two- term sets and plugging these solutions in to the remaining equation.  For instance, since term 5 must depend on $x$ ({\it cf} Table \ref{1termsets}), the set $\{1,2,3,6\}$ can only be irreducibly independent of $x$ if $\{4,5\}$ is irreducibly independent of $x$ as well.  However, from Table \ref{2termsets}, this would require $u=\frac{\alpha a^2}{2}+\gamma$.  Plugging this solution into  (\ref{Eqterms}) requires that $X_1 X_2 -\frac{dX_2}{dx}+\gamma X_1+aX_1^2+\frac{3\alpha a^2}{2}X_1$ be independent of $x$ in order for $\{1,2,3,6\}$ to be irreducibly independent of $x$.  However, as term 3 is the only term linear in $a$, the only solution is for $X_1$ to be constant, which leads to special case 1 and contradicts our assumptions.  Therefore,  $\{1,2,3,6\}$ cannot be irreducibly independent of $x$.

Finally, there are some four-term sets that cannot be eliminated in this manner.  For example, $\{1,2,5,6\}$ could be irreducibly independent of $x$ if a solution can be found that is compatible with the requirement that $X_1^2-\frac{dX_1}{dx}$ is constant which results from the two-term set $\{3,4\}$.  To check this possibility, we assume that $\{1,2,5,6\}$ is irreducibly independent of $x$.  In this case,
$X_1 X_2 - \frac{dX_2}{dx}+\frac{du}{da}X_2+\left(u+a\frac{du}{da}\right)X_1$ must be independent of $x$.  We do not know ${\it a ~priori}$ the functional form of $u(a)$ or $\frac{du}{da}$, but we do know that $\frac{du}{da}$ must depend on $a$ or this would reduce to a special case.  Since terms 1 and 2 are independent of $a$, the only way for this to be the case is if the $x$-dependence of $\frac{du}{da}X_2$ is cancelled by the $x$-dependence of $\left(u+a\frac{du}{da}\right)X_1$.  For this to be true, the $a$-dependence of $\frac{du}{da}$ must differ from the $a$-dependence of $u+a\frac{du}{da}$ by only a multiplicative constant, in which case
$X_2+\alpha X_1 = \beta$ for some constants $\alpha$ and $\beta$. By shifting the zero of $a$, we absorb ${X}_{2}$ into ${X}_{1}$ by shifting the zero of $a$.
So this particular case can be reduced to the special case of $X_2=0$ and does not yield any new solutions.

By examining all possible four-term sets, we find that no new solutions are admitted by any combinations that are not included as one of our special cases.  Applying the same method to five-term sets again yields no new results.  Thus, we have examined all possible combinations of one-, two-, three-, four-, five-, and six-term sets and discovered that they allow no solutions other than those covered by one of the special cases.  Since the special cases include all known conventional potentials, we therefore conclude that this method finds all of such known potentials, and it precludes other solutions.  We  have thus proven that the set of known $\hbar$-independent solutions is complete.

\section{Superpotentials that contain explicit $\hbar$ dependence}

Thus far we have found all known additive shape-invariant superpotentials that do not depend explicitly on $\hbar$, and have proven that no more can exist.
We now show that our formalism can be generalized to include ``extended" superpotentials that contain $\hbar$ explicitly.
In this case, we expand the superpotential $W$ in powers of $\hbar$:
\begin{eqnarray}
    W(x, a, \hbar) = \sum_{n=0}^\infty \hbar^n W_n(x,a)~. \label{W-hbar}
\end{eqnarray}
We wish to substitute (\ref{W-hbar}) in (\ref{SIC-SuperPotential}).
From  (\ref{W-hbar}),
\begin{eqnarray}\frac{\partial W}{\partial x}|_{a=a_0}=\sum_{n=0}^\infty \hbar^n \frac{\partial W_n(x,a_0)}{\partial x}~,\nonumber\end{eqnarray}
 and
\begin{eqnarray}W^2\left(x,a_0,\hbar\right)= \sum_{l=0}^\infty \sum_{k=0}^\infty \hbar^{k+l}W_kW_l.\nonumber\end{eqnarray}

Since $a_1=a_0+\hbar$,  $W\left(x,a_1,\hbar\right)= W\left(x,a_0+\hbar,\hbar\right).$  Expanding in powers of $\hbar,$

\begin{eqnarray}W\left(x,a_1,\hbar\right)=\sum_{m=0}^\infty \sum_{k=0}^{m}\frac{\hbar^m}{k!}\frac{\partial^k W_{m-k}}{\partial a^k}|_{a=a_0}.\nonumber\end{eqnarray}
So
\begin{eqnarray}W^2\left(x,a_1,\hbar\right)=
 \sum_{n=0}^{\infty}\sum_{s=0}^{n}\sum_{k=0}^{s}\frac{\hbar^n}{\left(n-s\right)!}\frac{\partial^{n-2}}{\partial a^{n-2}}\left(W_{k}W_{s-k}\right).\nonumber\end{eqnarray}
Similarly, 
\begin{eqnarray}\frac{\partial W}{\partial x}|_{a=a_1}=
\sum_{m=0}^{\infty}\sum_{k=0}^{m}\frac{\hbar^m}{k!}\frac{\partial^{k+1}W_{m-k}}{\partial a^k \partial x}.\nonumber\end{eqnarray}

After significant algebraic manipulation and requiring that the result
must hold for any value of $\hbar$, we find that the following equation must be true separately for each positive integer value of $n$:
\begin{eqnarray}\fl
\sum_{k=0}^n W_k\, W_{n-k} + \frac{\partial W_{n-1}}{\partial x}
-\sum_{s=0}^n \sum_{k=0}^s \frac{1}{(n-s)!}
\frac{\partial^{n-s}}{\partial a^{n-s}} W_k\, W_{s-k}
\nonumber \\
+\sum_{k=1}^{n} \frac{1}{(k-1)!}
\frac{\partial^{k}}{\partial a^{k-1} \,\partial x} \, W_{n-k}- \left( \frac1{n!}\,\frac{\partial^{n} g}{\partial a^n}   \right)=0.
\label{higherorders}\end{eqnarray}

For $n=1$, we obtain
\begin{eqnarray}
2\frac{\partial W_{0}}{\partial x}-\frac{\partial }{\partial a} \left(W_{0}^2+g  \right) = 0,
\label{lowestorder}\end{eqnarray}
yielding $2\frac{\partial^k W_{0}}{\partial a^{k-1}\partial x}= \frac{\partial^k }{\partial a^k} \left(W_{0}^2+g  \right)~{\rm for}~k\ge 1$. We have shown that all conventional superpotentials $W=W_0$ are solutions of this equation.  The extended cases (\cite{Quesne,odake}) are solutions to (\ref{higherorders}) as well.  In addition, new potentials can be generated by applying  (\ref{higherorders}) for all $n>1$, as we show in the next section.

\section{Generating a new $\hbar$ dependent superpotential}

We now use the method outlined in the previous section to generate a new potential.  We begin by choosing a conventional  shape invariant solution to satisfy (\ref{lowestorder}): $W_{0} = -a \tanh x - B {\rm sech} x$, which is a Scarf II superpotential from Table \ref{Table1} with $A=-a$.
For $n=2$, the expansion in (\ref{higherorders}) yields
\begin{eqnarray}
\frac{\partial W_{1}}{\partial x}-\frac{\partial }{\partial a} \left(W_{0}W_{1}  \right)
= 0~,\nonumber
\end{eqnarray}
and for $n=3$, we obtain
\begin{eqnarray}
\frac{\partial W_{2}}{\partial x}-\frac{\partial \left(2W_{0}W_{2}+W_{1}^2  \right)}{\partial a}
- \frac 12 \frac{\partial^2 W_{0}W_{1}}{\partial a^2}
+ \frac 23 \frac{\partial^3 W_0}{\partial a^2\partial x}  = 0.\nonumber
\end{eqnarray}
Without imposing boundary conditions, there are many solutions to (\ref{higherorders}) for each value of $n$.
However, from physical considerations we require that 1) solutions should not have singularities worse that $1/x^2$ to prevent
domain splitting or particles being sucked into the singularity; and 2) the asymptotic limits of $W$ be the same
as those for the corresponding $W_0$, so that supersymmetry remains unbroken.

With these considerations, these two coupled equations are solved by $W_{1}=0$ and
$W_2 = - \frac{B \cosh x}{(a-B \sinh x)^2}$.
The next order equations are solved by $W_{3}=0$ and $W_4 = - \frac{B \cosh x}{4(a-B \sinh x)^4}$.
Generalizing these, we get  $W_{0} = -a \tanh x - B {\rm sech} x$; and for $n\geq 1$, 
\begin{eqnarray}W_{2n-1}=0;~W_{2n} = - \frac{4 B \cosh x}{(2a-2B \sinh x)^{2n}},\nonumber\end{eqnarray}
yielding a sum that converges to
\begin{eqnarray}\fl
W(x,a,\hbar)=-a \tanh x-B{\rm sech}~x \nonumber \\
-2 B \hbar \cosh x\left(\frac{1}{2(a-B\sinh x)-\hbar} - \frac{1}{2(a-B\sinh x)+\hbar}\right).
\end{eqnarray}
This is a hitherto undiscovered superpotential that meets the requirements of shape-invariance.

\section{Conclusions}
We have transformed the condition for additive shape invariance into a set of local partial differential equations.    For conventional cases that do not depend on $\hbar$, we have shown that the shape invariance condition is equivalent to an Euler equation expressing momentum conservation for fluids and an equation of constraint.  Solving these equations we have generated all known $\hbar$-independent shape invariant superpotentials, and we have also shown that there are no others.

For extended cases in which the superpotential depends explicitly on $\hbar$, we  have developed an algorithm that is satisfied by all additive shape invariant superpotentials.  Finally, we have generated a new shape-invariant superpotential using this algorithm.  This method thus has the possibility to greatly expand our ability to identify shape-invariant superpotentials.

It may also be possible to extend this method to other forms of shape invariance such as multiplicative or cyclic. For these types of shape invariance, the potentials are generally not available in terms of known functions, except in very special cases ($N=2$ for cyclic and limiting cases for multiplicative). It remains to be shown whether for these,  the shape invariance condition can be transformed from a difference-differential equation into a set of partial differential equations and be subjected to similar analysis.

\begin{ack}
This research was supported by an award from Research Corporation for Science Advancement.
\end{ack}

\end{document}